\documentstyle[12pt]{article}
\input psfig.tex
\newcommand{\bFI}{\begin{figure}}
\newcommand{\eFI}{\end{figure}}
\newcommand{\bge}{\begin{equation}}
\newcommand{\ede}{\end{equation}}
\newcommand{\absz}[1]{\mbox{$\left |#1 \right |$}}

\newcommand{\Ha}{\begin{bf} H \end{bf} }

\begin{document}
\begin{center}
{\bf \Large Conductance in a periodically doped quantum wire}\\
J. Cserti, G. Sz\'alka and G. Vattay\\
Institute for Solid State Physics, \\
E\"otv\"os University, M\'uzeum krt, 6-8, H-1088 Budapest, Hungary
\end{center}

{\bf Abstract} - In this paper we will give a short rewiew about the conductance of a mesoscopic waveguide strip with a few impurities. 
Green-function method is used allowing to treat systems with low number of 
impurities, not only the completly clean case.
Investigating a wire containing Dirac-delta potentials (modelling the impurities) we found that 
increasing number of impurities can cause a transition of 
the structure of conductance. The staircase like structure of the clean 
system vanishes and the conductance of a system containing finite number of 
impurities located periodically will be determined by the band structure of the periodic system. 

\section{Introduction}

\begin{figure}[hbt]
\centerline{\strut\psfig{figure=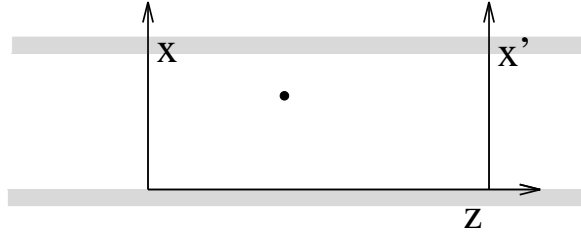,height=3cm}}
\caption{The lead with one impurity and the suitable coordinates \label{sys}}
\end{figure}

In recent years mesoscopic \cite{Beenakker} devices have been 
studyed intensively. Exact quantum mechanical calculations 
have been compared with short wavelength 
description of clean systems without impurity (ballastic regime), 
or with random matrix predictions for systems 
densely packed with impurities causing wild fluctuations of the potential 
on all length scales (diffusive regime). Nowadays thanks to the 
advances in manufacturing and material design one can get two 
dimensional electron gas in semiconductors whose behaviour is 
worth studying. These electrons can be considered as free electrons with 
effective mass. Using advanced nanotechnology waveguide strips with 
special geometry in extremely small size (50-500nm) can be produced. 
We investigated theoretically such kind of wire containing a few impurities (between 1 and 100) with the simpliest geometry we can imagine. (For one impurity see fig.\ \ref{sys}.)  

\begin{figure}
\centerline{\strut\psfig{figure=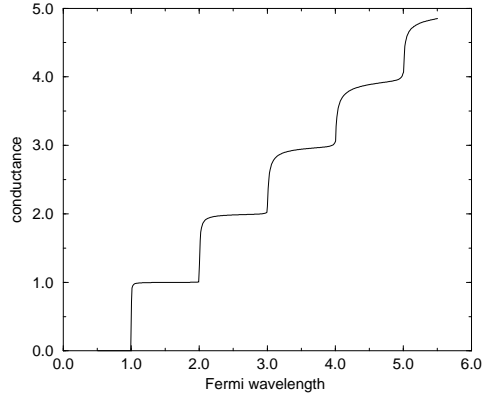,height=6cm}}
\caption{The conductance of the wire in case of one impurity in the universal 
units $2e^2/h$. The Fermi wavelength is plotted in units of $\pi/a$ where let $a$ denote the width of the wire. In this
unit a new channel always opens at the integers. \label{G1}}
\end{figure}

Investigating the conductance of this kind of systems experimentally gave a
surprising result. In this kind of extremely small systems 
in pure metals the conductance 
is a staircase like function of the Fermi wavenumber, 
with steps of heights  approximately $2e^2/h$. On Fig.\ \ref{G1}, where this 
behaviour can be seen, we show our 
result for the system on Fig.\ \ref{sys}. The explanation of 
this kind of behavior was given first by Landauer and B\\"uttiker \cite{Landauer}.

Similar studies can be found in Ref.\ \cite{EgyDelta,sajatcikk,TDK} for the case of one impurity modelled by Dirac-delta potential.
 
In this paper we first explain the Landauer formula which one may use
for calculating the transmission or reflexion matrixelements of a system. 
Then in  section ``Green-function method'' we give
a short review of Green-function method which is very helpful treating 
systems containing low number of impurities. We then  consider 
the conductance of a periodic system (see Fig.\ \ref{peri}) and at the end of the last section we compare conductance calculated in section \ref{Gfth} with Green-function method and the one
for the periodic case developed in section ``Periodic case''.

\section{Landauer formula}

The conductance of the 
non-degenerate electron gas in a waveguide considered here is given 
by the Landauer formula \cite{Landauer,BarSto}. According to this 
theory, incoming and outgoing wave functions in the leads far from 
impurities can be decomposed into incoming and outgoing 
quantum modes
\begin{eqnarray}  
\psi_n(x,z)&=&\Phi_n(x)e^{ik_nz},\nonumber \\  
\Phi_n(x)&=&\sqrt{\frac{2}{a}}\sin(\frac{\pi n x}{a}), 
\end{eqnarray}    
where $k_n=\sqrt{2mE_F-(\hbar\pi n/a)^2}/\hbar$ is the wave number of 
the propagating planewave, $a$ is the width of the lead and $E_F$ is 
the Fermi energy.  If the Fermi energy is less than 
$E_n=\frac{1}{2m}(\hbar\pi n/a)^2$ the 
wavenumber $k_n$ becomes imaginary and the $n^{th}$ channel becomes closed 
preventing wave propagation.

The Landauer formula for the conductance $G(E_F)$ at Fermi
energy $E_F$ is 
\begin{equation}  
G(E_F)=\frac{2e^2}{h}\sum_{m,n} |t_{mn}|^2, 
\end{equation}  
where $t_{nm}$ is the transmission probability amplitude from the 
incoming channel $n$ on the entrance side to the outgoing channel $m$ 
on the exit side.  In case of infinitely long leads summation goes for 
open channels of both sides only.  The transmission probability is given by
the projection of the Green-function over the transverse
wave functions $\Phi_n(x)$ on the entrance lead for the incoming modes
and $\Phi_m(x')$ on the exit lead for outgoing modes
\begin{equation}  
t_{nm}=2i(k_nk_m)^{1/2}\int dxdx'\Phi_n(x)G^+(r,r',E)\Phi_m(x'), 
\label{trans} 
\end{equation}
where $r,r'$ denote the vector $(x,z),\,(x',z')$ and  
$z$, $z'$ can lie anywhere on the entrance side and  exit side, 
respectively.
Reflection between two 
modes $n$ and 
$m$ on the same side is given by 
\begin{equation}  
r_{nm}=\delta_{nm}-2i(k_nk_m)^{1/2}\int dxdx'\Phi_n(x)G^+(r,r',E)\Phi_m(x'), 
\label{refl} 
\end{equation}  
where $z$ lies anywhere on the entrance side.  For open channels $n$ of the 
entrance side the transmission and reflection amplitudes fulfill the sum 
rules 
\begin{equation}  
\sum_m |t_{nm}|^2+\sum_{m'}|r_{nm'}|^2=1, 
\label{check}
\end{equation}  
where the two summations go for channels $m$ on the exit side and $m'$ 
on the entrance side. These are the consequences of the current conservation.

\section{Green-function method \label{Gfth}}

The Hamiltonian of the system with one impurity can be separated 
into two parts
\bge 
\Ha=\Ha_{0}+\Ha_{1}.
\label{h0}
\ede
where 
\bge
\Ha_{0}=-{\Delta}
\ede
is the free particle Hamiltonian with effective mass $m$ of the electron and
\bge
\Ha_{1}(r'')={\lambda}{\delta}(r''-r_{0})
\ede
models the potential of the impurity, (see Fig.\ \ref{sys}.) using $\hbar=1,m=\frac{1}{2}$.

In order to study the case with impurity the Dyson equation can be used.
It gives as the Green-function of the system we get adding a new potential ({\bf V}) to a system whose Green-function is known $(G_0)$
\bge 
G(r,\,r',\,E)^+=G_{0}^+(r,\,r',\,E)+\int dr''G_{0}^+(r',\,r'',\,E){\bf V}(r'')G^+(r,\,r'',\,E).
\label{dys}
\ede 
If we know the Green-function of the empty waveguide strip ($G_{0}$) and we add one Dirac-delta potential at $r_1$ with the strength of $\lambda$, the Green-function of the wire with one impurity will be the following:

\bge
\label{iter}
G^+(r,\,r',\,E)=G_0^+(r,\,r'\,E)+{\lambda} \frac
{G_0^+(r,\,r_{1},\,E)G_{0}^+(r_{1},\,r',\,E)}
{1-{\lambda}G_{0}^+(r_{1},\,r_{1},\,E)}.
\ede

Including more impurities we can apply the Dyson equation more times for 
calculating the Green-function for this case \cite{Grosche}. 

In our method first the Green-function of the empty waveguide needs to be 
calculated. In our case this Green-function can be defined as

\bge
 (E-{\Ha}_0)G_0^+(r,\,r',\,E) = {\delta}(r-r'). 
\label{defGreen}
\ede 
The solution of this equation (\ref{defGreen}) is  

\bge
G_{0}^+(r,\,r',\,E)=\sum_{n}\frac{{\Phi}_{n}(x){\Phi}_{n}(x')}{2ik_n}
e^{ik_n\absz{z-z'}}\;.
\label{G_0}
\ede
where
\bge
k_n=\sqrt{E-\frac{n^2}{a^2}{\pi}^{2}}.
\label{kndef}
\ede

Once we know 
the Green-function using (\ref{trans}) the conductance can be developed. 
The results we can get using the tools developed here will be 
considered at the end of the next section.

\section{Periodic case \label{pepe}}

\begin{figure}
\centerline{\strut\psfig{figure=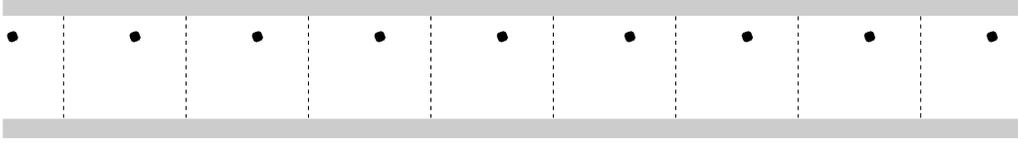,height=1.9cm}}
\caption{The periodic case, where the location of the impurity is x=0.86, z=0.65, the length of the blocks are 1 and the width of the lead, a is also 1. \label{peri}}
\end{figure}

In this section we calculate the electronic band structure 
of a periodic system, where each unit cell contains one impurity 
(see Fig.\ \ref{peri}). 
Treating the periodic case we have the following Hamiltonian

\bge
\frac{{\hat p}^2}{2m}\Psi(x,z)+V(x,z)\Psi(x,y)=E\Psi(x,y).
\label{hm}
\ede
It is wellknown that for periodic system the wavefunction can be given as

\bge
\Psi_{k,n}=e^{ikz}u_{k,n}(x,y),
\label{pszi}
\ede
where $u_{k,n}(x,z)$ is yet unknown function, and $k$ is a Bloch number.
Inserting (\ref{pszi}) into (\ref{hm})
 we can obtain a new kind of Hamiltonian for 
the function of $u_{k,n}(x,z)$.
\bge
\frac{1}{2m}\bigg{(}\hbar k+\frac{\hbar}{i}\nabla 
\bigg{)}^2u_{k,n}(x,z)+V(x)u_{k,n}(x,z)=E_{k,n}u_{k,n}(x,y),
\label{hami}
\ede
At the first step of the calculation, the Green-function of the empty system
will be developed. In this case the $V(x,z)$ potential is zero, except the
hard wall boundary of the wire, which corresponds to 
Dirichlet boundary condition.
It is easy to obtain the solution that satisfies the equation (\ref{hami}), 
with the Dirichlet boundary condition.
\bge
u_{k,n}(x,z)=\sqrt{\frac{2}{a}}sin(\frac{n\pi x}{a}),\;\;\;n=1,2,...
\label{hull}
\ede
Note that it does not depend on $z$. The eigen-energies of the empty wire are given then as
\bge
E_{k,n}=\frac{\hbar^2}{2m}\bigg{(}(k+K)^2+\frac{n^2\pi^2}{a^2}\bigg{)},
\ede
where $K=2\pi j/L$ belongs to the reciprocal lattice (L is the length of the unit cell, and j denotes any integer).The Green-function of the empty wire 
is 

\bge
G_{0,k}(r,r',E)=\frac{2}{a}{\sum_K}{\sum_n}\frac{\sin(\frac{n\pi x}
{a})\sin(\frac{n\pi y}{a})}{E-E_{k,n}}.
\ede
Now using Dyson equation the Green-function of the system each block 
containing one Dirac-delta potential at $r_0$ with the strength $\lambda$ can 
be written as

\bge
G_k(r,r',E)=G_{0,k}(r,r',E)+\lambda\frac{G_{0,k}(r,r_{0},E)G_{0,k}
(r_{0},r',E)}{1-\lambda G_{0,k}(r_{0},r_{0},E)}.
\label{dy}
\ede
In order to find the eigen-energies of this system with impurities, one has to find the singularities (poles) of the Green-function (\ref{dy}).
A short calculation can show that
the only singularity we have here comes from the denominator of the formula (\ref{dy}). Then the equation for the eigen-energies will be the following

\bge
0=1-\lambda G_{0,k}(r_{0},r_{0},E(k))\,,
\ede
which one has to solve and the eigen-energies ($E(k)$) can be calculated. 

\begin{figure}
\centerline{\strut\psfig{figure=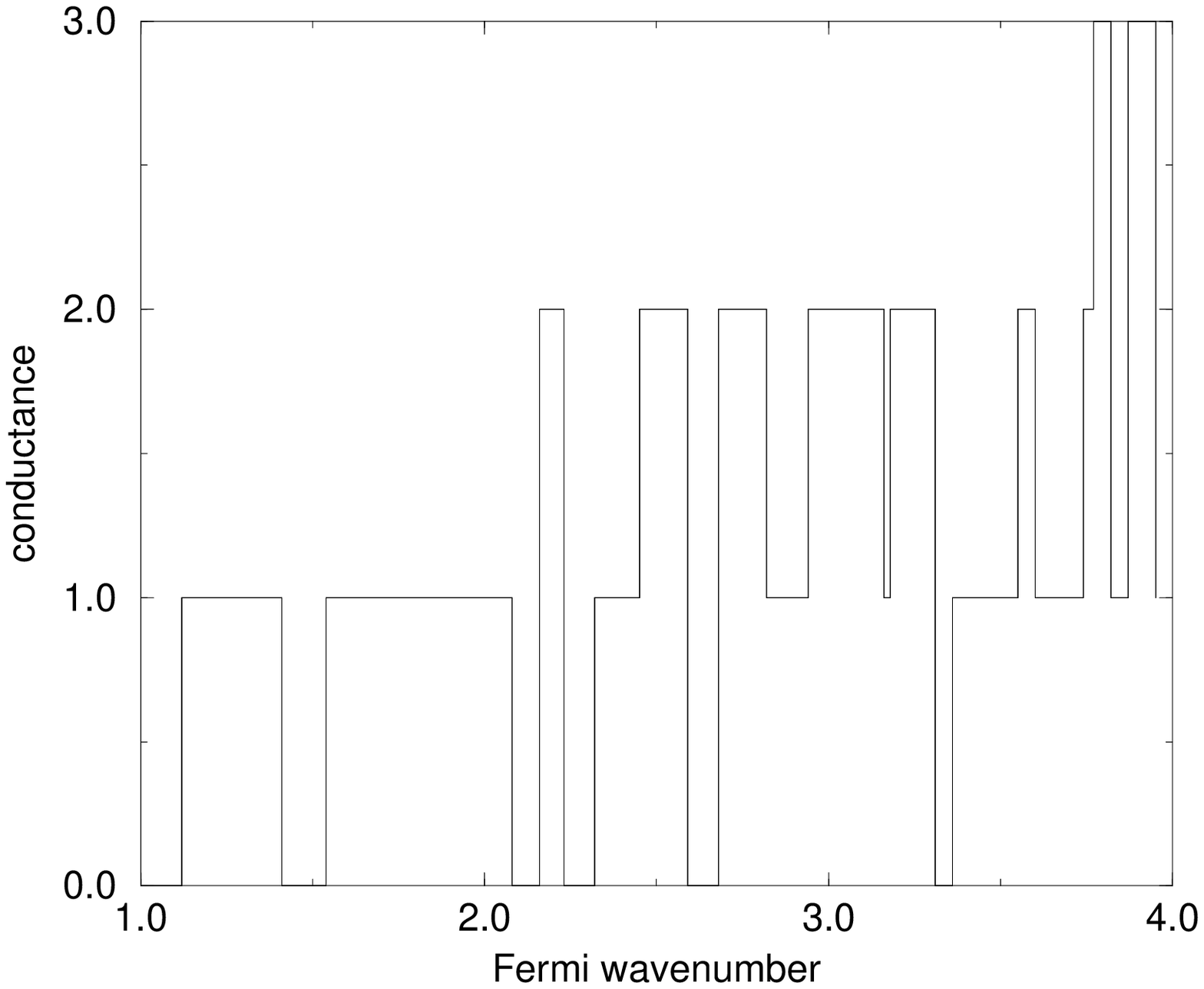,height=7cm}}
\centerline{\strut\psfig{figure=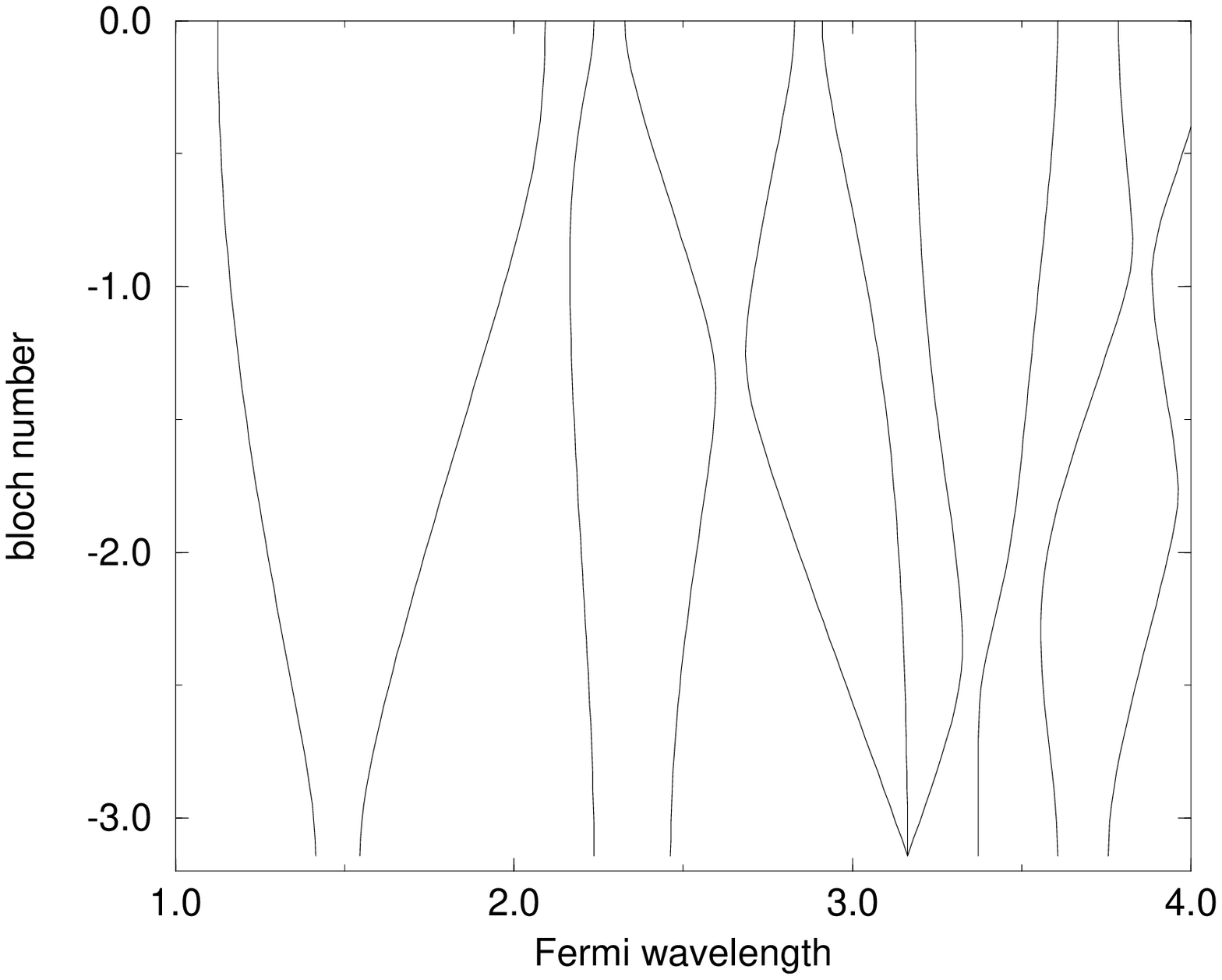,height=7cm}}
\caption{The bandstructure of the system (below) and the conductance of the periodic case calculated from the given bandstructure (above). The units as usuall.\label{id}}
\end{figure}

As soon as  we have the band structure $(E(k))$ of the wire (Fig.\ \ref{id}. below), one may calculate the conductance of the periodic system. It is wellknown, that for calculating the conductance, the number of bands at a given energy level has to be counted. Each band gives a unit of ($2e^2/h$) to the conductance. 

\begin{figure}
\centerline{\strut\psfig{figure=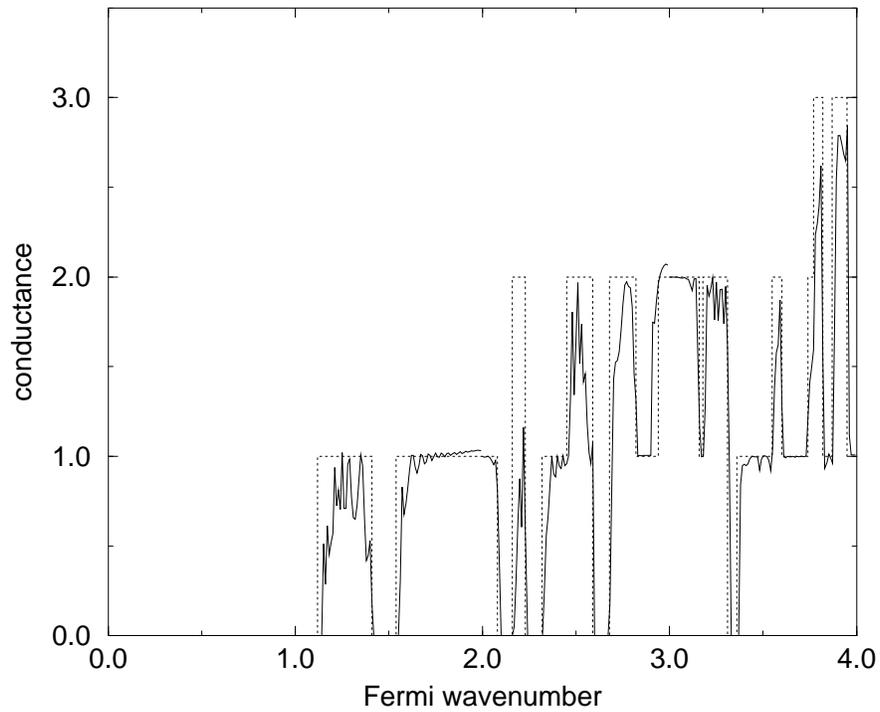,height=11cm}}
\caption{We can compare the two kind of conductance. The dotted 
line denotes the shape of the periodic case calculated from the given band 
structure (see also fig.\ \ref{id}). The continuos line denote the 
conductance of the system where 50 blocks where connected and also 
the two extremely long empty lead \label{cond}}. 
\end{figure}

Now let us use the results of the Green-function method (section \ref{Gfth}). 
We have seen the conductance for the case of one single Dirac-delta. 
This case can be regarded as a block which contains the potential, 
connected to two extremely (infinitely) long empty wires. The periodic case can be built up from the same blocks. Now the next step is to connect two such 
\begin{figure}[hbt]
\centerline{\strut\psfig{figure=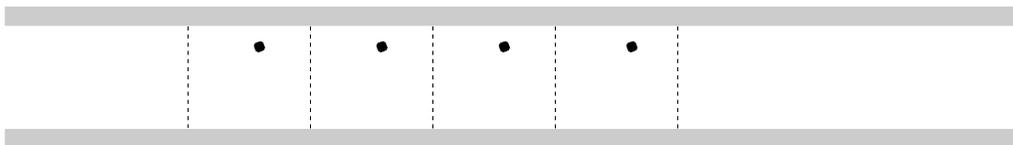,height=1.9cm}}
\caption{Low number of impurities in a lead. Each block are equivalent, however it is {\em not} the periodic case. In the blocks the location of the impurity is x=0.86, z=0.65, the length of the blocks are 1 and the width of the lead, a is also 1. \label{veges}}
\end{figure} 
blocks, and then the two empty wires. We can go on with this iteration, 
we can connect any number of the {\em same} blocks and the two infinitely long wires.
The case of four blocks can be seen on Fig.\ \ref{veges}. Using Landauer 
formula and the Green-function method we can easily calculate the 
conductance of these systems. The results show a very interesting behaviour. 
If there are a few blocks in the wire (1-3) the staircase structure seen on 
Fig.\ \ref{G1} remains structurally unchanged with some deformation of the shapes of the steps. However inserting 
more impurities a new situation occurs. We can hardly see the reminescences of 
the structure of the Fig.\ \ref{G1}, a new structure is developing. 
Once we compare this kind of conductances with the conductance calculated 
from the 
periodic system, we can see that between these two kinds of 
conductance, the difference is diminishing with increasing number of 
the inserted blocks (Fig.\ \ref{cond}).

\section{Summary}

In this paper a short review has been given about the transport behaviour of mesoscopic leads
using Green-function theory. Thanks to the Dirac-delta potential an exact 
Green-function is available for the case of low number of impurities. As soon as
the band structure is known, the conductance of the periodic case 
is also known. When the number of the impurities in the lead are 
increasing, the ideal staircase steps are deforming, and a new structure, 
determined by the band structure of the periodic system occurs.

\section{Acknowledgments}

We would like to thank S. Witoszynskyj the special opportunity to write 
this paper, P. Pollner, A. Csord\'as, T. Tasn\'adi, G. Tichy their interest, and encouragement. This work has been supported by the Hungarian Science Foundation 
OTKA (F019266/F17166/T17493), Hungarian Ministry of Culture and Education (FKFP0159/1997), and the Israeli-Hungarian collaboration
grant (OMFB-ISR9/96).


\end{document}